# Explore Spatiotemporal and Demographic Characteristics of Human Mobility via Twitter: A Case Study of Chicago

Feixiong Luo · Guofeng Cao* · Kevin Mulligan · Xiang Li

**Abstract** Characterizing human mobility patterns is essential for understanding human behaviors and the interactions with socioeconomic and natural environment, and plays a critical role in public health, urban planning, transportation engineering and related fields. With the widespread of location-aware mobile devices and continuing advancement of Web 2.0 technologies, location-based social media (LBSM) have been gaining widespread popularity in the past few years. With an access to locations of hundreds of million users, profiles and the contents of the social media posts, the LBSM data provided a novel modality of data source for human mobility study. By exploiting the explicit location footprints and mining the latent demographic information implied in the LBSM data, the purpose of this paper is to investigate the spatiotemporal characteristics of human mobility with a particular focus on the impact of demography. To serve this purpose, we first collect geo-tagged Twitter feeds posted in the conterminous United States area, and organize the collection of feeds using the concept of space-time trajectory corresponding to each Twitter user. Commonly human mobility measures, including detected home and activity centers, are derived for each user trajectory. We then select a subset of Twitter users that have detected home locations in the city of Chicago as a case study, and apply name analysis to the names provided in user profiles to learn the implicit demographic information of Twitter users, including race/ethnicity, gender and age. Finally we explore the spatiotemporal distribution and mobility characteristics of Chicago Twitter users, and investigate the demographic impact by comparing the differences across three demographic dimensions (race/ethnicity, gender and age). We found that, although the human mobility measures of different demographic groups generally follow the generic laws (e.g., power law distribution), the demographic information, particular the race/ethnicity group, significantly affects the urban human mobility patterns.

**Keywords:** Human mobility, geodemography, social media, Twitter, name analysis, Chicago

Feixiong Luo · Guofeng Cao · Kevin Mulligan
E-mail: {feixiong.luo, guofeng.cao, kevin.mulligan}@ttu.edu
Department of Geosciences, Texas Tech University

Xiang Li
E-mail: xli@geo.ecnu.edu.cn
Key Lab of Geographical Information Science, East China Normal University

Corresponding author: Guofeng Cao (guofeng.cao@ttu.edu)

## 1 Introduction

Human mobility plays an important role in the study of traffic forecasting (Kitamura et al. 2000), disease spreading (Colizza et al. 2007), urban planning (Horner & O'Kelly 2001), and in the general science and engineering of smart cities (Batty 2012). Conventionally, human mobility research heavily relies on traditional surveys of travel journals as major data sources. For example, based on 30 travel surveys in more than 10 countries, Schafer (2000) has shown that human travels exhibit strong regularities across space and time. The process of conventional survey is typically long, expensive, and often limited to a small number of samples, and hence difficult to realistically deal with the complex human behaviors. The advancement of Web and mobile technologies yields a set of reliable and cost-effective data sources that can provide fine-granularity of spatiotemporal information for large scale human behaviors and social dynamics. Typical examples of these data sources include on-line bank note tracking logs (Brockmann et al. 2006), mobile phone calling records (Gonzalez et al. 2008), vehicle GPS trajectories (Yuan et al. 2012), and transactions of bank (or credit) cards (Sobolevsky et al. 2014, Lenormand et al. 2015) and transportation cards (Hasan, Schneider, Ukkusuri & González 2013). By taking advantage of these data sources, numerous findings, such as spatiotemporal regularity and heavy-tail distribution of travel distances, have been recently reported (Barabasi 2005, Yan et al. 2011).

Most recently, social media (e.g., Twitter and Facebook), a set of on-line applications that allows users to create and exchange contents, have been experiencing a spectacular rise in popularity and attracting hundreds of millions of users for social networking, content generating and sharing. With the wide adoption of location-aware mobile smart devices and wireless communications, people tend to access social media from mobile smart devices, and thanks to the in-built positioning capability, location and whereabouts information can be attached to the social media messages (e.g., geo-tagged Twitter posts, Flickr photos and check-ins). The location-based social media (LBSM) data provide access to the locations as well as the contents of the social media activities, hence provide a promising modality of data source for studying the complex human behaviors and understanding socioeconomic dynamics (Liu et al. 2015). By capitalizing upon this new data source, novel and successful applications have started to emerge. In geography, location-based social media have been used to spatiotemporally and schematically characterize the geographic places (McKenzie & Janowicz 2015, McKenzie et al. 2015). Specifically in human mobility research, by analyzing 22 million check-in records in multiple location sharing services, Cheng et al. (2011) demonstrated that human mobility is a mixture of short, and random movements with occasional long jumps. Cho et al. (2011) found that social network relationships can explain approximately 10% and 30% of human movements and 50% to 70% periodic behaviors based on check-in records and friendship networks from LBSM sites Gowalla and Brightkite. Hasan, Zhan & Ukkusuri (2013) explored an archive of Foursquare and Twitter posts, and reported that users tend to visit different urban places with diminishing regularity governed by a Zipf's law. Based on a collection of almost a billion tweets recorded in the year of 2012, Hawelka et al. (2014) demonstrated the characteristics of international travels and of human mobility across different countries. Wu et al. (2014) combined activity-based analysis with a movement-based approach to model the intra-urban human mobility observed from about 15 million social media check-in records in China.

Most of the above mentioned research focuses on the spatial and temporal aspects of the human mobility and tends to ignore the effects of demographic and other (e.g., socioeconomic and health status) background information that have been known as significant factors in population distribution and human mobility. Exceptions include Cheng et al. (2011) and Li et al. (2013), which linked LBSM activities with associated geographic regions (e.g., via spatial join) to explore differences of LBSM activities as local geographic and economic landscapes vary at aggregated scales. Compared with the previously mentioned data sources (e.g., cell phone calling logs, bank notes) that have only access to the user locations, LBSM data also offer access to the profiles (e.g., profile names) of individual social media users and contents of social media messages. These information, if analyzed appropriately, can provide important background (e.g., demographic information and health status) of the social media users. These learned background information, together with location footprints, makes it possible to investigate the spatiotemporal and demographic characteristics of human mobility at a fine individual scale.

Personal names (both first and last names) have been demonstrated to carry rich amount of demography (Mateos 2007), geography (King & Jobling 2009), cultural (Zelinsky 1970), linguistic and even genetic information (King & Jobling 2009) about the name bearers, and name analysis has been used to study a range of problems in different disciplinary, e.g., ethnicity and population structure (Mateos & Longley 2011), human migration (Piazza et al. 1986). Twitter requires new users to provide a full name (a pair of forename and surnames). Although real full names are not enforced, a significant fraction of active social media users are willing to provide real names (Peddinti et al. 2014), which have been successfully used to help demography breakdowns of social media users (e.g. Gallagher & Chen 2008, Oktay et al. 2014, Chen et al. 2013). In this paper, we apply a similar name analysis to the names provided in Twitter profiles to detect demographic groups, and then investigate the impact of the demographic background on spatiotemporal distribution and human mobility of Twitter users. Specifically, based on a public data stream of Twitter feeds posted in the conterminous United States area, we first represent the location footprints of each Twitter user in terms of a space-time trajectory, and for each of the trajectories, commonly used human mobility measures, including detected home locations, activity centers and radius of gyration, are computed and maintained. A subset of Twitter users whose detected home locations in the city boundary of Chicago are chosen for a case study. Name analysis is applied to the Chicago Twitter users to detect the implicit groups of three demographic factors, including race/ethnicity, age and gender. We then investigate the spatiotemporal distribution and mobility characteristics of Chicago Twitter users and compare the differences across the detected groups of the three demographic factors.

The remainder of this paper is structured as follows. Section 2 introduces the overall work flow of this paper including the methodology for detecting demographic information and activity centers of Twitter users. Section 3 implements the methodology, analyzes the spatiotemporal and demographic characteristics of human mobility of Chicago Twitter users. Section 4 concludes the paper and discusses the limitations and future work.

## 2 Methodology

Figure 1 gives the overall flowchart of the proposed analysis of this paper. Five consecutive steps (components) are included: (1) Twitter feed collection, (2) space-time trajectories construction, (3) local residents identification, (4) name analysis for demographic groups, and finally (5) investigating spatiotemporal and demographic characteristics of human mobility of Chicago Twitter users.

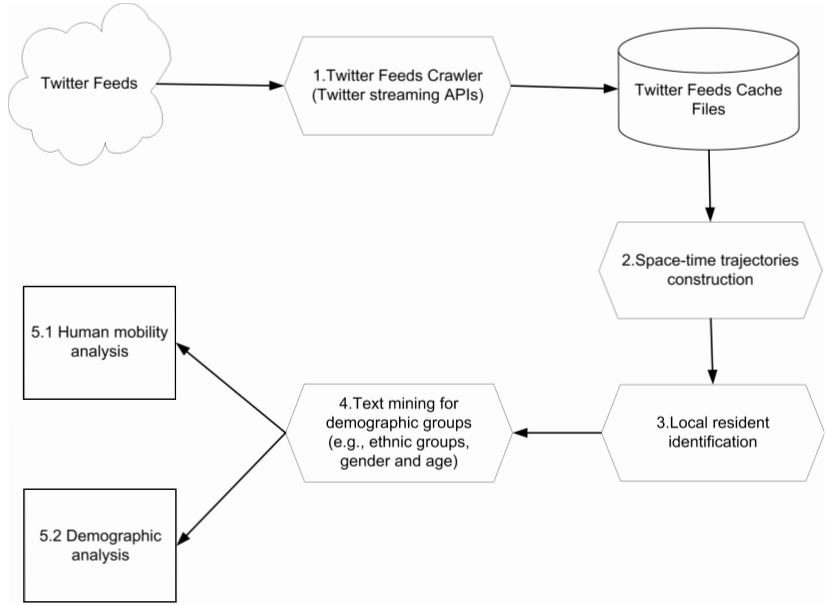

**Fig. 1** The flowchart of the proposed method.

The first step is to retrieve data from Twitter. A tweet crawler is developed based on the Twitter streaming API (application programming interface) to collect tweets posted in the contiguous United States area. We collected six months of geo-tagged tweets between January 1, 2013 and June 30, 2013. The dataset consists of over 300 million records generated by a total of over 3 million users. Each geo-tagged tweet has contents of the tweets and the associated exact location and timestamp, and each Twitter user has a unique identifier and a profile name. Despite the 1% limit of sampling, it has been reported that the streaming API returns almost the complete set of the geo-tagged tweets that are of interest of this paper (Morstatter et al. 2013). Based on the collections of Twitter posts, the following steps were conducted in order to better proximate human mobility via geo-tagged tweets.

## 2.1 Space-time trajectories

The raw collections of Twitter posts are first organized in terms of space-time trajectories, each corresponding to an individual Twitter user. The concept of space-time trajectories (Hägerstrand 1970) has been well recognized as a simple and effective mean for visualizing and exploring a person's activity trajectory, and been successfully used in characterizing human mobility pattern (e.g., Miller 1991, 1999, Kwan 1998, 2000). We assume that each Twitter user $u_{id}(id \in [1, N])$ corresponds to a continuously moving, space-time trajectory $T_{id}$ in a geographic space (illustrated as a blue solid line in Figure 2). The "true" trajectory $T_{id}$ is measured by a time series of social media activities (illustrated as *red* points in Figure 2), $S_{id} = \{(\boldsymbol{p}_{id}^0, t_{id}^0, m_{id}^0), (\boldsymbol{p}_{id}^1, t_{id}^1, m_{id}^1), \dots, (\boldsymbol{p}_{id}^i, t_{id}^i, m_{id}^i), \dots\}$, where each tuple $(\boldsymbol{p}_{id}^i, t_{id}^i, m_{id}^i)$ denotes the footprint of user $u_{id}$ who applies the social media activity $m_{id}^i$ at location $\boldsymbol{p}_{id}^i \in \mathcal{R}^2$ and time-stamp $t_{id}^i \in \mathcal{R}$. Different from traditional trajectories of moving objects (Zheng & Zhou 2011), the space-time trajectory in location-based social media data for an individual user could be very sparse (Gao & Liu 2013) and episodic (Andrienko et al. 2012). It is usually difficult to estimate the activities and locations between samples $S_{id}$. In this paper, we assume that a social media user $u_{id}$ stays at the same location during $[t_i, t+1]$ as at $t_i$ until a new activity is posted at $t_{i+1}$ when $u_{id}$ moves from $\boldsymbol{p}_i$ to $\boldsymbol{p}_{i+1}$ (illustrated as a red line in Figure 2).

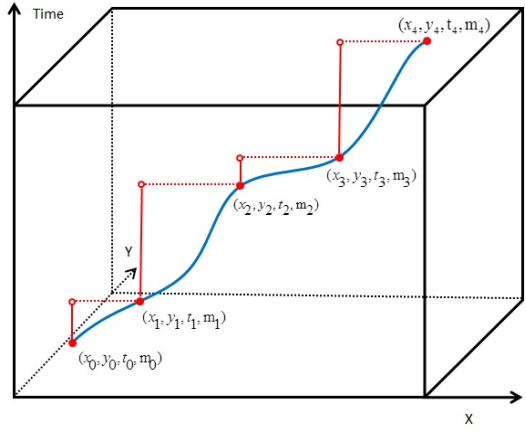

**Fig. 2** An illustration of space-time path constructed from location-based social media data in a space-time cube of Hägerstrand (1970).

In the resulting database of trajectories, trajectories with less than 24 tweets during the half year of period (on average a tweet per week) were filtered out because the tweets are too sparse to proximate individual movement profiles. To characterize the massive number of space-time trajectories associated with social media users, two widely used geometric measures (e.g., in Gonzalez et al. 2008, Cheng et al. 2011, Hasan, Schneider, Ukkusuri & González 2013), *radius of gyration* and *activity centers* including home locations, are computed and updated for each space-time trajectory.

### 2.1.1 Radius of gyration

The concept of radius of gyration or gyradius, originally for describing the mass distribution of an object, has been adopted to characterize the activity range of trajectories (Gonzalez et al. 2008). It amounts to the standard deviation of distances between points on a trajectory and the center point. A low value of radius of gyration indicates the associated user travels mainly locally, while a high value indicates the user have long-distance travels. The radius of gyration for a user $u_{id}$ can be formalized as:

$$r_{id} = \sqrt{\frac{1}{N_{id}} \sum_{i=1}^{N_{id}} (\boldsymbol{p}_{id}^i - \overline{\boldsymbol{p}_{id}})^2}$$

where $N_{id}$ is the size of the trajectories samples in $TS_{id}$ (the number of this user's social media activities) and $\boldsymbol{p}_{id}^i - \overline{\boldsymbol{p}_{id}}$ is the distance between each sample and the geographical center of the trajectories $\overline{\boldsymbol{p}_{id}} = \frac{1}{N_{id}(t)} \sum_{i=1}^{N_{id}(t)} \boldsymbol{p}_{id}^i$.

### 2.1.2 Activity centers

An activity center represents a place or location that an individual frequently visits for a certain purpose. For example, most of people usually rest at home, work in their offices and eat at restaurants and etc. We refer to the locations of these frequently visited places (home, office, and restaurants) as activity centers. One may have more than one centers for a specific type of activity, and similarly as the radius of gyration, activities centers could help characterize the human mobility and life pattern. It should be expected that, for users with regular life patterns, the number of activities centers of tend to converge to a stable number over the time, while for active users, it might be not the case. The activity centers of a user $u_{id}$ could be inferred from the locations of trajectory samples $TS_{id}$. In this paper, a well-known spatial clustering algorithm, namely DBSCAN (density-based spatial clustering of applications with noise) algorithm (Ester et al. 1996), is adopted to determine activity centers of users. We refer the readers to Ester et al. (1996) for detail procedure of the DBSCAN algorithms. Without presuming the number of activity centers, this algorithm has shown advantages over other supervised methods in clustering spatial and spatiotemporal points, and has been widely used across different fields (e.g., Huang et al. 2014). It should be noted that an activity center is a spatial concept and subjective to the effects of geographic (spatial) scales. Depending on the scenarios of study, the scale of an activity center could range from a city to a point of interests (e.g., a coffee shop).

### 2.1.3 Detection of home locations

Among the activity centers, the location of home plays a particularly important role in understanding demographic characteristics of human and human behaviors, and is also crucial to distinguish between residents and visitors of a place. Similar with many existing works (Cheng et al. 2011, Huang et al. 2014, e.g.,), the "home" in this paper is defined as the most frequently visiting place for a user during night-time. Specifically, the following steps are used to determine their detected homes:

(1) For each social media user, we first filter out the social media activities posted from 20:00 to 8:00 am, and then apply DBSCAN method to the locations of the selected activities to determine activity centers.
(2) The most frequently visited activity centers during 20:00 to 8:00 am is chosen as this user's detected home.

## 2.2 Surname analysis for race/ethnicity

Knowing the home of the Twitter users not only helps distinguish visitors from residents, also sheds light on the sociodemographic characteristics of the residents because the sociodemographic profiles of a community is correlated with those of the residents, and the data of the former are often publicly available. For example, knowing a person lives in a census block with high percentage of African American population provides information for estimating that person's race due to the effect of racial segregation.

In addition to the home location, a person's name (a pair of forename and surname) bears rich amount of information about the person. Surnames provide clues of geographic, linguistic, social, ethno-cultural and demographic background (Mateos & Longley 2011), while the forenames provide more hints on gender and ages. Twitter requires new users to provide full names to use the social media services. Although real names are not enforced and in many cases meaningless tokens or pseudonyms are given, a significant fraction of active social media users are willing to provide real names (Peddinti et al. 2014) and the provided names are included in Twitter streaming API that is used to collect the Twitter posts. By analyzing the names in Twitter user profiles, together with the sociodemographic profile of the community that they reside, we can reveal more demographic information (e.g., ethnic group, gender and age) of Twitter users that are important to understand population distribution and human mobility characteristics. In this paper, we analyze the surnames to detect the associated race/ethnicity groups and analyze the forenames for age and gender information.

To detect the race/ethnicity group associated with a surname, we first linked (spatially joined) the detected home address (represented in a pair of latitude and longitude) of Twitter users to the associated US census tracts. Then we adopted a Bayesian method, the so-called Bayesian Improved Surname Geocoding method (BISG) (Elliott et al. 2009), to integrate the race/ethnic information implied in the surnames and the sociodemographic profiles of the census tracts where the Twitter users reside. We focus on the five major race/ethnicity group in the US: (1) White, (2) African-American, (3) Hispanic, (4) Asian people and (5) Others. Specifically, for the surname dataset, the US Census Bureau released a detail list of $151,671$ surnames (shared by more than 100 persons), along with each surname's self-reported race/ethnicity population distribution. In other words, each surname is associated with a vector of population numbers or prior probabilities (say $P_r$) for each race/ethnicity that the name bearer should be. For the demographic information of census tracts, we use the demographic profile datasets from the decennial census of 2010, which consists of the total population and proportion of each race/ethnic group and demographic segment. Hence for an individual with a race/ethnicity group, the census data can provide the probability that the individual reside in this census track (say $P_c$). Given a surname, the BISG can then integrate the two probabilities ($P_r$ and $P_c$) through a Bayesian approach to estimate the posterior probability of the surname belongs to each racial/ethnic group. And the race/ethnicity with the max possibility value is considered as this user's race/ethnicity. By using multiple sources of information, the BISG tends to yield higher estimation accuracy than other existing methods (Elliott et al. 2009), such as the Categorical Surname and Geocoding approach (Fiscella & Fremont 2006) and the Bayesian Surname and Geocoding method (Elliott et al. 2008). The readers are referred to Elliott et al. (2009) for details of the BISG method.

### 2.3 Forename analysis for gender and age

As surnames carry information on ethnic background, forenames are more gender-specific and since the forename popularity tends to evolve over the time, thus provide a strong prior for the gender and age of the name bearer. Similar as the databases for surnames in the US, databases of forenames are available. US Social Security released a forename database (US Social Security Administration 2014), containing $104,110$ forenames with more than 5 occurrences and associated gender, year of birth since 1880 till February of 2015. The database has $39,199$ male names and $64,911$ female names, and $10,221$ overlapping for both males and female. The names in the database are from Social Security card applications for births that occurred in the United States. Tang et al. (2011) generated a similar name datasets with $23,363$ names by crawling Facebook profile pages and removing fake names. Similar as the US Social Security database, each name in the

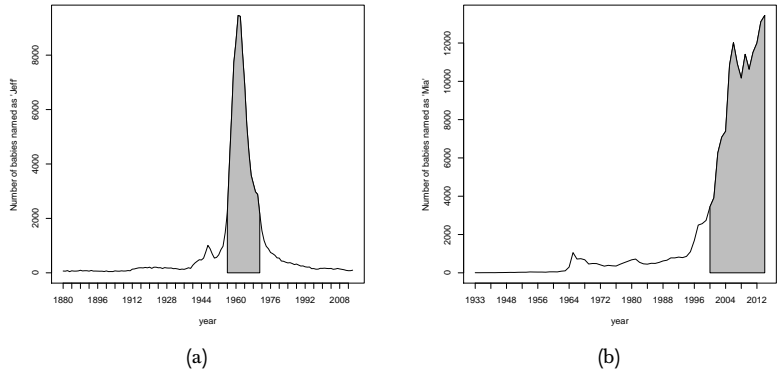

(a)                                                  (b)

**Fig. 3** Temporal trends of the numbers of new born with forename as "Jeff" (a) and "Mia" (b) from the year of 1880 to 2014: 73% of persons named Jeff were born during the year of 1956 - 1971 and 83% of Mia were born after the year of 2000.

Facebook-based list has the number of occurrences that the name was labeled as male and female. This Facebook-based forename dataset has been reported to achieve 96.3% accuracy of gender estimation when applying on Facebook (Tang et al. 2011). Considering the high overlapping percentage (90% of Twitter users are also Facebook users) between Twitter and Facebook users (Pew Research Center 2013), we used the Facebook-based database for gender estimation. Specifically, given a forename in our collected name pairs, if it appears in the Facebook-based name list, we first calculated the gender probability based on the fraction of occurrences that were labeled as male and female, then assign this name the gender with higher probability.

This Facebook-based database, however, does not contain age or year of birth information for each name. Hence we use US Social security name dataset (US Social Security Administration 2014) to estimate the age groups of users. This database has been successfully used for estimation of age, gender and even identification (Gallagher & Chen 2008, Chen et al. 2013, Oktay et al. 2014). To illustrate the idea of using forenames for age estimation, Figure 3 displays the popularity changes of two popular names, "Jeff" and "Mia". Both of these two names show obvious temporal trend in the occurrences; 73% of persons named "Jeff" were born during the year of 1956 - 1971 (gray region) and 83% of "Mia" were born after the year of 2000 (the gray region). Hence we can estimate the rough age groups according to the temporal trend. Specifically, we are interested in three age groups: (1) $\leq 20$ (2) $21 - 60$ and (3) $\geq 61$ based on the assumptions that for most people between 21 and 60 is roughly the working age (Bureau of Labor Statistics 2014).

(1) All the first names in the US Social Security baby name database are grouped into these three groups according to the year of birth. For example, "Lucy" is a popular baby name in 2000. It is assigned into the first group ($\leq 20$) according to the above classification. Each group was assigned to a empirical probability based on the fractions of number of occurrences.

(2) Each first name in our user name-pairs are compared against the US Social Security name database. If matched and the maximum probability is larger than a specified thresh-hold value, the age group with maximum probability is assigned to this name.

## 3 Analysis and results

We applied the previously discussed methodology to the collections of Twitter post. Specifically for the city of Chicago, we first identified the local residents of Chicago, and extracted the tweets posted only by Chicago local residents. Please note that Chicago residents could post tweets outside of Chicago. The resulting Chicago datasets includes $7,967$ Chicago local users and corresponding space-time trajectories consisting about 2 million tweets across the US, among which 85.7% were posted in the Chicago boundary and 14.3% were outside. We then applied name analysis to estimate the race/ethnicity group, gender and age for each Twitter user. Based on these analysis, this section first explores the spatiotemporal distribution of the total Chicago Twitter users, and then from the three demographic dimensions, decompose the Twitter users into different segments and analyzes the mobility characteristics respectively.

### 3.1 Spatiotemporal characteristics of tweeting activities

The spatial distribution of Chicago Twitter local users is shown in Figure 4(b), where each dot represents the detected home location of a Twitter user. It can be seen that user density in the north of the city is higher than the south, and especially the north area next to the Lake Michigan has the highest density. The spatial distribution of the detected homes of Twitter users closely represents the distribution of Chicago population of 2010 decennial survey at a census tract level (Figure 4(a)). This can be further verified by the linear correlation (with correlation coefficient as 0.44) between the number of the detected residents out of the Twitter users in each census tract and the associated censused population shown in Figure 4(c).

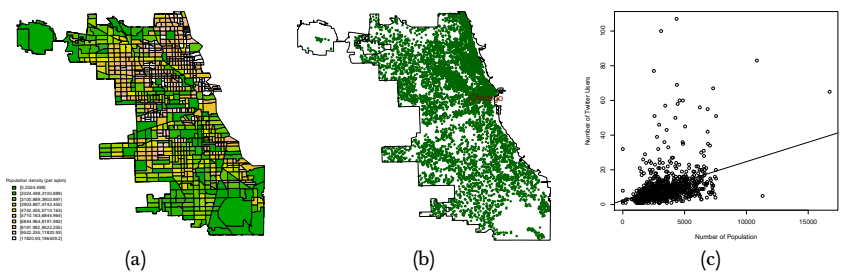

(a)                    (b)                    (c)

**Fig. 4** (a) The spatial distribution of Chicago population in census 2010 at a census track level; (b) The spatial distribution of home of Chicago Twitter users; (c) The relationship between the number of Twitter users and the censused population in Chicago.

The overall spatiotemporal pattern of tweeting activities by Chicago local users is shown in Figure 5. The daily average of tweets posted by Chicago local users is about $20,000$ (Figure 5(a)), amounting to each Twitter user averagely posting about 3 tweets per day. Figure 5(a) shows a spike of tweet count on $17th$ March, 2013. This could be because this date is the *St. Patrick's day*, a traditional large event in Chicago. Figure 5(b) demonstrates the overall temporal pattern of tweeting activities at each day by week (Monday to Sunday), and obviously, all days share similar temporal patterns with slight

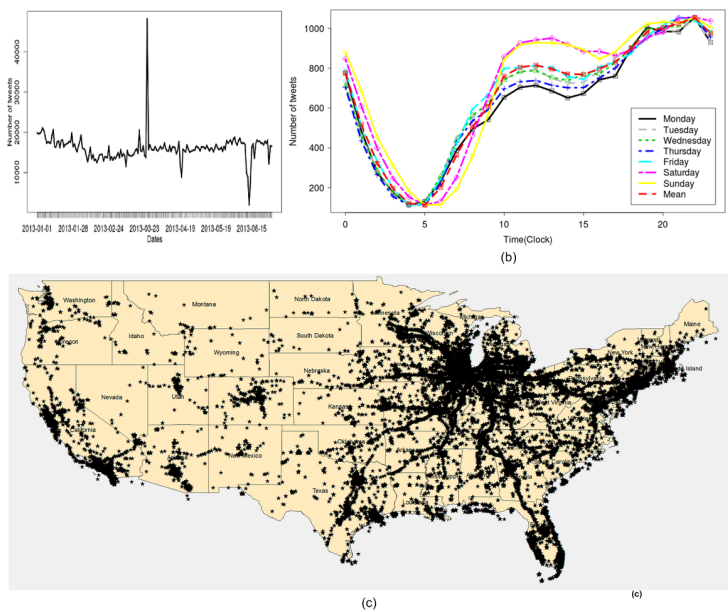

**Fig. 5** (a) The daily average number, (b) hourly average number, and (c) spatial distribution of geo-tagged tweets posted by Chicago local Twitter users.

differences for weekends. The highest rate of tweeting occurs approximately from $10:00$ in the morning to $22:00$ at midnight. There are generally two tweeting peaks: one around $13:00-14:00$ and the other around $20:00-21:00$. The lowest rate of tweeting is roughly $4:00-5:00$ in the morning. Working days from Monday to Friday have a slightly different pattern from weekends. The tweeting rate in the weekend is clearly higher than that of working days approximately from $11:00$ in the morning to $16:00$, and lower from $5:00$ to $9:00$ in the morning. Figure 5(c) shows the Twitter activities of local Chicago residents across the US. It is easy to see that that Chicago and its neighboring area have highest density of tweets. The tweets density of the central and northeast of US is higher than that in southern and western of US, and spatially, the tweet density aligns well with major corridors through Chicago. Although not approximate to Chicago, major cities with large population (e.g., Los Angeles, San Francisco, Denver, Dallas, Houston, Phoenix) also stand prominently in Figure 5(c).

### 3.2 Spatiotemporal characteristics of human mobility

Radius of gyration (or gyradius) and activity centers are two commonly used metrics of human mobility. The distribution of gyradius is shown in Figure 6 (a) with mean and median values as 183.03 and 20.56 kilometers respectively. It can be observed that its distribution follows the power law, consistent with the previous findings such as Gonzalez et al. (2008). Similar as Jurdak et al. (2015), the distribution could be approximately divided into three segments in terms of the distribution line slope. The first segment is the gyradius values from 0 kilometer to about 25 kilometer with slope as -0.02 (illustrated as the red line in Figure 6), corresponding to the local Chicago scale; the second segment is the gyradius values from 25 kilometers to 1000 kilometers with the slope as -0.99 (illustrated as the green line in Figure 6) (a), corresponding to the regional scale of Chicago surroundings; the third segment is the gyradius values from 1000 kilometers to 2000 kilometers with the slope as -8.87 (illustrated in the blue line in Figure 6) (a), corresponding to the national scale. Obviously, the first segment has the highest density and the slope stays relatively steady meaning the majority of Twitter users travel locally within Chicago during the period of study and share the similar gyradius values. Since urban human mobility pattern is constrained by spatial structure of a city, the distribution of the gyradius values in this segment reflects the urban structure and morphology of the Chicago city (Kang et al. 2012). In the second and the third segments, the density starts to decrease and decrease faster as the gyradius values increases, meaning fewer Twitter users tend to travel actively over longer distances than local Chicago scales and much fewer actively travel across the nation.

The spatial distribution of the gyradius values of users is shown in Figure 7(a). Same as in Figure 4 (b), each point represents the detected home location of a Twitter user and the color represents the values of gyration radius. It can be observed that the majority of users with large gyradius value tend to live in the downtown and the north communities that are next to the Lake Michigan. The temporal trend of radius of gyration is shown in the Figure 8(a). It can be easily seen that the radius of gyration increases from 70.61 kilometers in January to 183.04 kilometers in June and tends to converge over the time (Figure 8(a)).

Similarly, the histogram of the number of activity center is shown in Figure 9(a) with mean and median values as 5.4 and 4 respectively. Over 50% of people have 2-4 activity centers and less than 10% people have more than 10 activity centers. The population fraction gradually decreases with the increase of the number of activity centers. The spatial distribution of the number of activity centers of Twitter users is shown in Figure 7(b). Different from that of the gyradius, Figure 7(b) does not show significant spatial patterns. The temporal pattern of the number of activity centers is shown in Figure 8(b). Each user has averagely 1.4 activity centers in the first month (January), and is accumulated to 5.4 in the June. Unlike the convergence trend in the distribution of gyration radius (Figure 8(a)), the number of activity center increases steadily without obvious trend of convergence.

Spatially, Figure 10(a-f) show the density maps of Chicago activity centers for each month (January to June, accumulatively). These maps were generated using kernel density estimation (based on a Gaussian kernel with bandwidth as 500 meters) (Silverman 1986) provided by the `spatstat` package (Baddeley et al. 2004) in R computing environment (Gentleman et al. 1997). Not surprisingly, this sequence of maps show that the hotspots in Downtown area, the most active area of Chicago with the highest density of points of interests (e.g., offices, restaurants, and recreational parks), stand out first (Figure 10(a-b)), then the hotspots keep spreading to the north and west to the downtown (Figure 10(c-f)). The spatial patterns in the map sequences start to converge since April. Similar diffusion and convergence patterns were also found beyond the Chicago city boundary across the United States: hotspot areas first appear in the city of Chicago and its neighboring areas and then gradually expands into surrounding Midwest areas, and then major cities of other farther regions. As in Figure 5, this should reflect the spatial patterns of socioeconomic links of the Chicago city with the rest regions of the US.

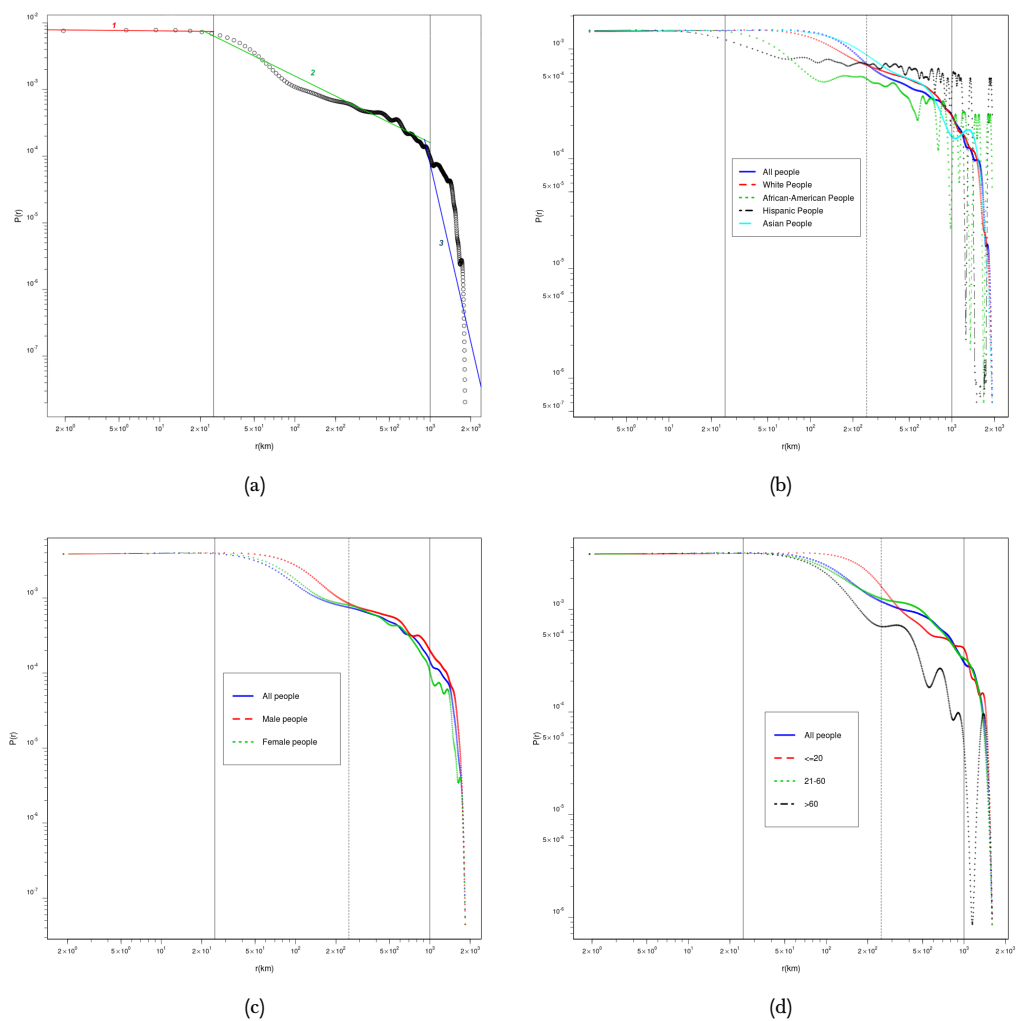

(a)                                                          (b)

(c)                                                          (d)

**Fig. 6** The density distribution of radius of gyration of (a) all the Chicago Twitter users, (b) Chicago Twitter users with detected races, (c) Chicago Twitter users with detected gender; (d) Chicago Twitter users with detected age groups.

## 3.3 Demography characteristics on human mobility

The demographic information of the Chicago Twitter users was determined using the name analysis methods discussed in Section 2. The results are summarized in Table 1. Specifically:

(l) Races/ethnicity information of 3,010 users, approximately 38% of the total users, were successfully identified. These include 2,228 (74%) White, 261 (8.7%) African-American, 461 (15.3%) Hispanic and 60 (2.0%) Asian people. This race/ethnicity breakdown for Twitter users closely represents the findings of Oktay et al. (2014).

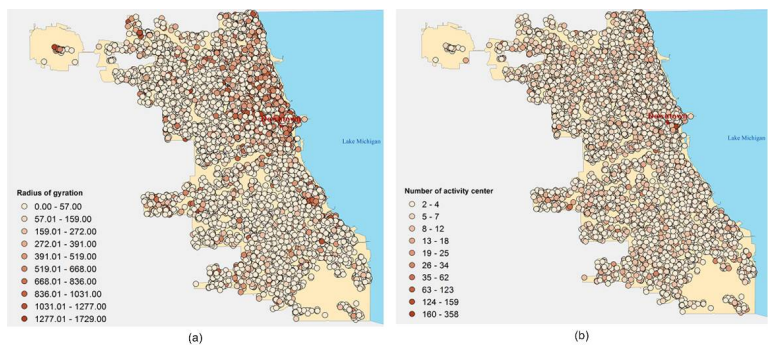

**Fig. 7** (a) The spatial distribution of the radius of gyration of Chicago Twitter users; (b) The spatial distribution of the number of activity center of Chicago Twitter users.

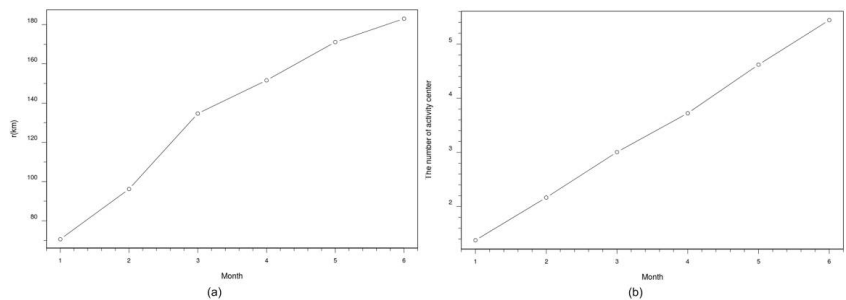

**Fig. 8** (a) The average radius of gyration over month; (b) The average number of activity centers over month.

(2) Gender information of $4,301$ users, approximately 54.0% of the total users, were successfully identified. Among them includes $2,280$ (53%) male and $2,021$ (47%) female. This gender breakdown for Twitter users matches well with the online users survey of Pew Research Center (2015).

(3) Similarly, age information of $1,123$ users, approximately 14.1% of the total users, were successfully identified. Among them includes 355 (31.6%) no more than 20 year old, 710 (63.2%) between 21 and 60 year old, and 58 (5.2%) greater than 60 years old. Again, this age group breakdown for Twitter users closely represents the findings of Oktay et al. (2014).

The detected home locations of users with identified races, gender and age groups is shown in Figure 11. Each point represents the location of a user's detected home. Figure 11 (a) shows apparent spatial separations among residential areas of different detected races/ethnicity groups, while the detected gender and age groups (Figure 11 (b-c)) do not show this obvious pattern. The detected African-American group tends to live in the southern and western neighborhoods of the city and demonstrates more segregated pattern than the detected White, Asian and Hispanic users, which tend to mixed with each other. This pattern matches well with previous findings (e.g., Massey & Denton 1989) and the real distribution of people of different race/ethnicity groups in Chicago (Massey & Denton 1989, Weldon Cooper Center for Public Service 2014).

The density distribution of gyradius values for different races is shown in Figure 6(b), where the blue dashed line is for all the users with successfully identified race (with mean and median values as 232.83 and 73.82 kilometers respectively), red for White (with mean and median as 275.56 and 140.33 km), green for African-American (with mean and median as 85.13 and 7.97 km), black for Hispanic (with mean and median as 84.89 and 7.2 km) and cyan for Asian (with mean and median as 313.92 and 87.68 km). It can be seen that Twitter users with each identified race/ethnicity group follow similar power law distribution as the total Chicago Twitter users (Figure 6(a)). The gyradius values of White and Asian people tend to be similar and larger than those of African-American and Hispanic people. Same as in Figure 6(a), the distribution can be decomposed into three segments with the first segment from 0 km to 25 km, the second from 25 km to 1000 km and the third from 1000 km to 2000 km. It can be observed that in the first segment the density distribution of radius of gyration of people of different races including White, African-American, Hispanic and Asian people follows roughly the same trend. However, in the second segment, the situation is not that straightforward. The density values for the detected Hispanic people tends to be flat than other three groups. There is a turning point at 250 kilometers (indicated by a dash vertical line in Figure 6). Before 250 kilometers, the probability density of gyradius values of White and Asian people are higher than those of Hispanic and African-American people. After 250 kilometers, the density values for Hispanic people jumps to the highest, then White and Asian people, and the density for the African-American people stays the lowest. The distribution of the African-American and Hispanic people becomes unstable in the third segments, which could be due to the lack of enough users whose gyradius values are from 1000 km to 2000 km.

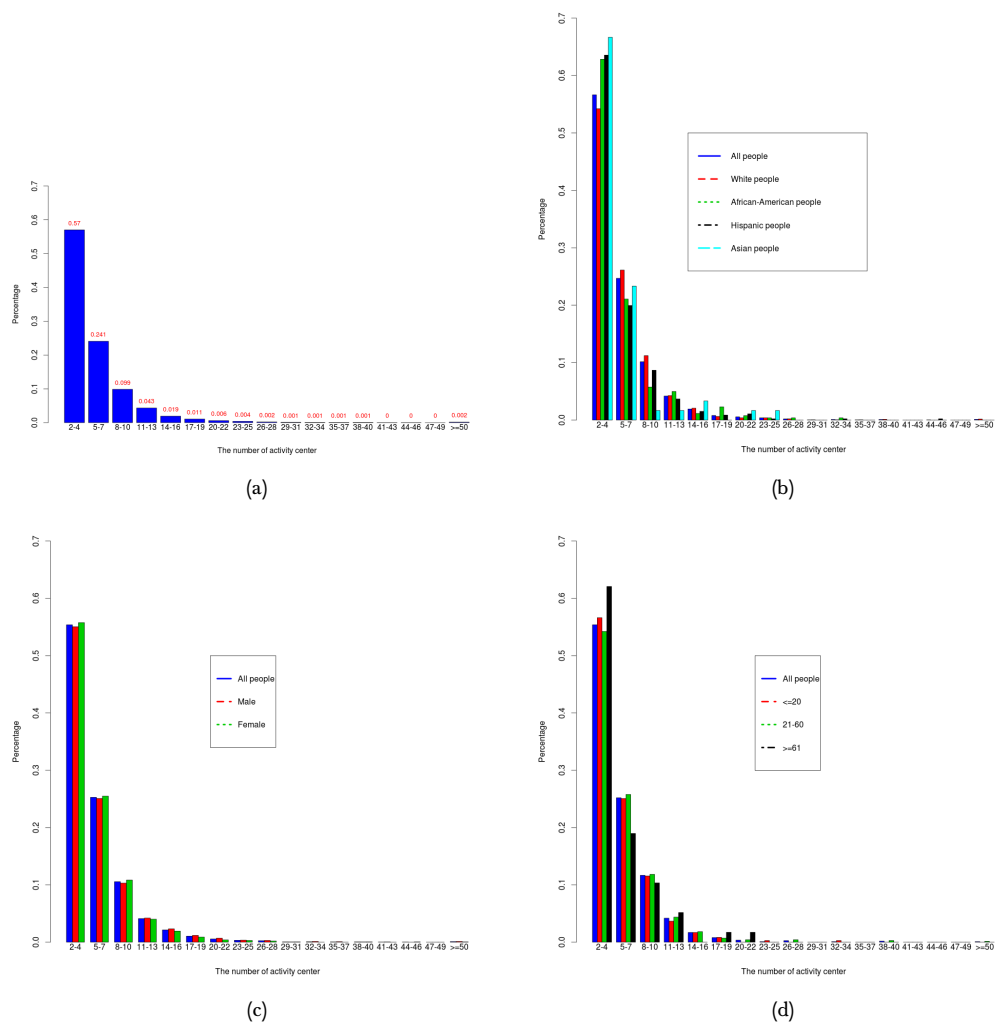

**Fig. 9** The density distribution of the number of activity centers of (a) all the Chicago Twitter users, (b) Chicago Twitter users with detected races, (c) Chicago Twitter users with detected gender; (d) Chicago Twitter users with detected age groups.

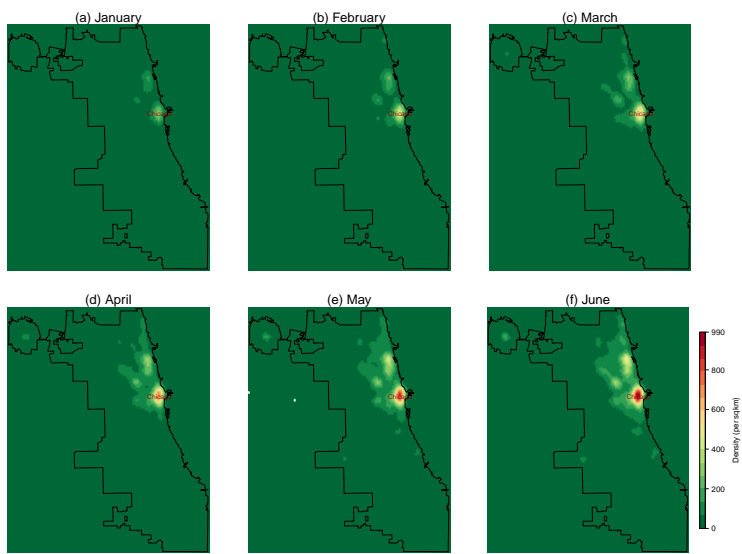

**Fig. 10** The spatial distribution of activity centers of Chicago Twitter users in Chicago over the 6 months.

| Number of total users | Number of users (unidentified race) | Number of users (identified race) | | | | |
|---|---|---|---|---|---|---|
| 7967 | 4957 | Total | White | African-American | Hispanic | Asian |
| | | 3010 | 2228 | 261 | 461 | 60 |
| | Number of users (unidentified gender) | Number of users (identified gender) | | | | |
| | 3666 | Total | Male | Female | | |
| | | 4301 | 2280 | 2021 | | |
| | Number of users (unidentified age) | Number of users (identified age) | | | | |
| | 6844 | Total | ≤ 20 | 21-60 | >60 | |
| | | 1123 | 355 | 710 | 58 | |

**Table 1** The breakdown of demographic groups of Chicago Twitter users

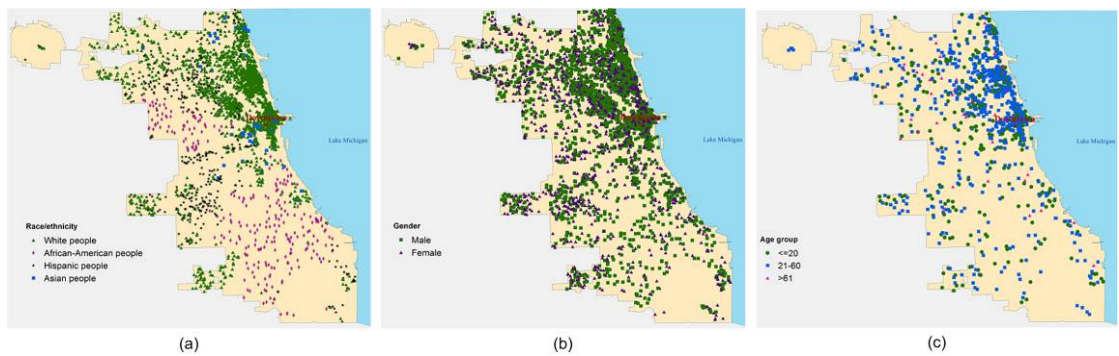

**Fig. 11** The detected home locations of Chicago Twitter users with different (a) detected races, (b) detected gender; (b) detected age groups.

The density distribution of gyradius values for people with different gender is shown in Figure 6(c), where the blue dashed line is for all the users with successfully identified gender (with mean and median values as 233.711 and 67.69 km), red for male (with mean and median as 244.07 and 67.62 km) and green for female (with mean and median as 222.02 and 68.45 km). Similarly, the distribution of radius of gyration of male people and female are both consistent with the power law distribution of all people in Figure 6(a). Moreover, the distribution can be divided in the same way as in Figure 6(b). Specifically, the probability density of male and female people are very close in the first and third segments. The density of male people is a higher than that of female people in the first section of the second part while their densities are close in the latter section of the second part. Similar density plot is shown for people with different age groups in Figure 6(d), where the blue dashed line is for all the users with successfully identified age groups (with mean and median respectively 247.19 and 84.23 km), red for 20 years old and younger (with mean and median 207.93 and 45 km), cyan for between 21 and 60 years old (with mean and median as 274 and 119.69 km) and black for 61 years and the older (with mean and median as 159.02 and 18.7 km). Similar power law trends are observed for different age groups. Again, the plot is divided in the same way as in Figure 6(b). The probability density for different age groups is almost the same in the first segment. The density values for more than 60 years old people is lowest in the second and third segments (the unstable zigzag pattern is due to the lack of enough data points). The density values for no more than 20 years old people is slightly higher than that of people between 21 and 60 years in the first section of the second segment and this trend becomes opposite in the latter section of the second part. And in the third segment, the density values for no more than 20 years old people and between 21 and 60 years people is fairly close.

We also conducted the comparative analysis for the number of activity centers. The distribution of the number of activity centers for Twitter users with different detected races is shown in Figure 9 (b). The blue bars represent the total users with detected race/ethnicity information, the red bars are for White, green for African-American and cyan for Asian. It can be observed that although the activity number distribution of different identified race/ethnicity have similar overall pattern with that of all users in Figure 9 (a), differences exist among people of four races. Specifically, White users tend to have the highest percentage with median size of centers (5-10 centers) and lowest percentage with small size of centers (2-4 centers). African-American people have almost have the highest percentage with the relatively large size of centers (11-19) and the lowest percentage for median size of centers (5-10 centers). Hispanic people do not have an obvious distribution characteristic at all intervals. Asian people have a rather sporadic distribution with the highest percentage with small size of centers (2-4 centers) and large size of centers (14-22 centers). The distribution of the number of activity centers for users with detected gender are shown in Figure 9 (c). Male and female users share similar power law distribution with averagely 5 activity centers. Similarly for different age groups, the distribution of the activity center numbers are show in Figure 9 (d). It can be observed that the distribution of more than 60 years old is slightly different from the other two age groups. Specifically, more than 60 years old tend to have the higher percentage of with small size of the number of centers (2-4 activity centers) and the large size of the number of activity center (17-22 activity centers), and the other two groups share similar distribution.

Spatially, the density distribution of activity centers of users with different races in Chicago is shown in Figure 12 (a-d). Similar with Figure 10, these kernel density maps are generated using Gaussian kernel with 500 meters bandwidth. The four race/ethnicity groups demonstrate different spatial distribution patterns while sharing a common hotspot area at the downtown of Chicago. White and Asian users are more spatially concentrated than the detected African-American and Hispanic users. Similarly as the detected home locations in Figure 11(a), the activity center of the detected White users tend to have high density in the north to the downtown next to the Lake Michigan in addition to the downtown area (Figure 12(a)), African-American users in the south and west (Figure 12(b)), Hispanic users in southwest and northwest (Figure 12(c)) and Asian users in the dowtown proximity and north to the downtown (Figure 12(d)). It can be also observed that, similar as their detected home locations (Figure 11(a)), the activity centers of the detected African-American users tend to be more spatially seperated from the other three other groups (Figure 12 (a-d)). Figure 13 shows the spatial distribution of activity centers of Chicago Twitters users with different races across the US. These kernel density maps for the contiguous US areas are generated using a Guassian kernel with 20 kilometers as bandwidth in the Contiguous United States Albers Equal Area projection. Hotspots of White users' activity centers tends to spread across the whole US area, while those of the other three groups tend to concentrate in Chicago area with fewer spots in major large cities. In addition to the common hotspots in surrounding areas of Chicago and large cities like New York and Los Angeles, interestingly for African-American users, hotspots also appear in the southern major cities in the states of Mississippi, Alabama and Georgia (e.g., Atlanta, Houston, Memphis and Birmingham), and for Hispanic users, Dallas and San Antonio in Texas, and Miami and Orlando in Forida, appear as hotspots. As in Figure 5 (c), this could reflect the socioeconomic connections of Chicago with these cities for different race/ethnicity groups. Figures 14 and 15 show no apparent differences for spatial distribution of male and female users' activity centers. For the three age groups, one can see from Figures 16 and 17 that user groups of no more than 20 years old and 21-60 years share similar similar patterns, while the group of 60 year older have more concentrated activity area in Chicago and fewer hotspots outside of Chicago than the other two age groups.

## 4 Discussion and conclusion

In this paper, we explored the spatiotemporal and demographic characteristics of urban human mobility using a collection of geo-tagged Twitter posts. Based on the detected home location of each anonymous Twitter user, we first selected the Twitter users that reside in the city of Chicago as a case study. Radius of gyration and activity centers were used to describe the urban human mobility of Chicago Twitter users. Similar as previous findings based on other sources of data, e.g., cell phone calling logs, transportation or credit card transactions, the human mobility measures generally follow the power law distribution. Temporally, the radius of gyration of an individual tends to converge over the time, while the number of activity centers do not show such trend for an individual. Spatially, the distribution of activity centers within Chicago boundary aligns well with the socioeconomic development of the city, and the activity center distribution out of Chicago reflects the socioeconomic links of Chicago with the areas.

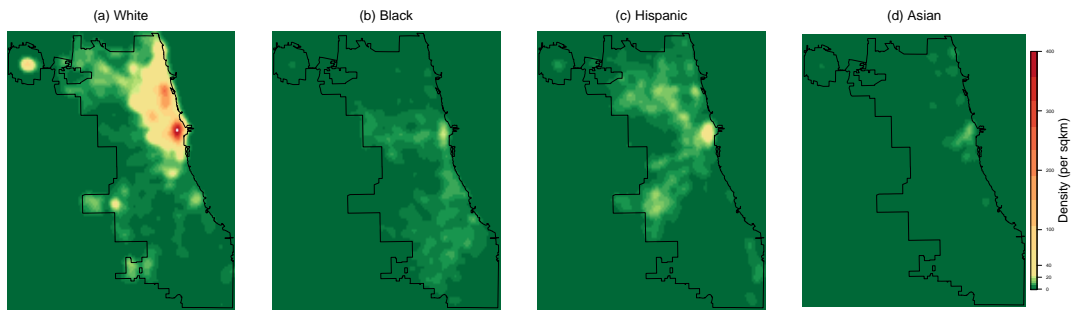

**Fig**. **12** The spatial distribution of activity centers in the Chicago area of Chicago Twitter users with different detected race/ethnicity groups.

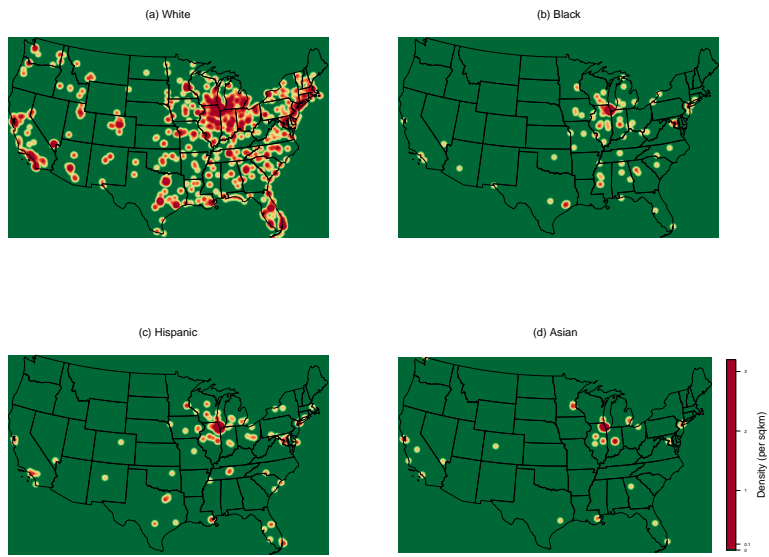

**Fig**. **13** The spatial distribution of activity centers in the US area of Chicago Twitter users with different detected race/ethnicity groups.

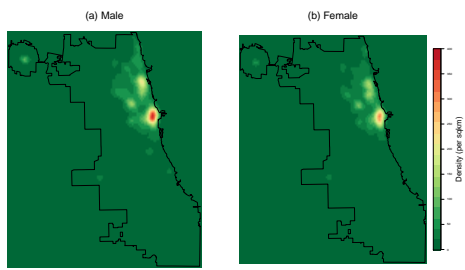

**Fig**. **14** The spatial distribution of activity centers in the Chicago area of Chicago Twitter users with different detected gender.

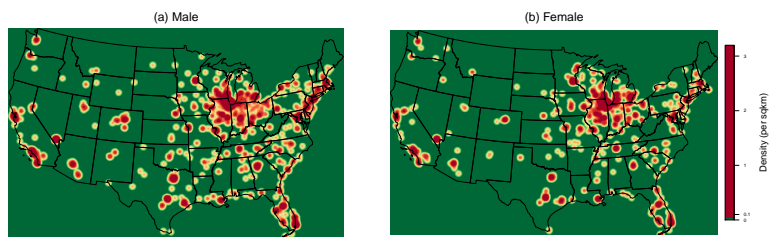

**Fig. 15** The spatial distribution of activity centers in the US area of Chicago Twitter users with different detected gender.

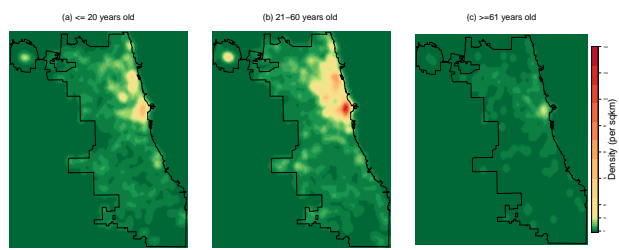

**Fig. 16** The spatial distribution of activity centers in the Chicago area of Chicago Twitter users with different detected age groups.

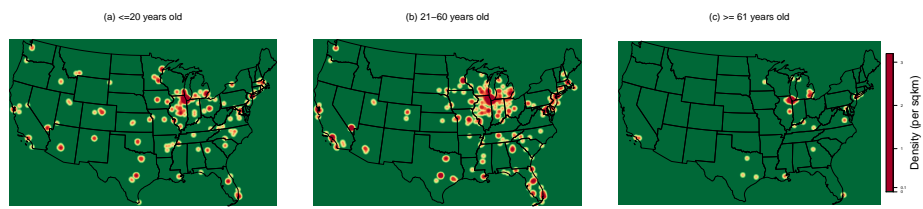

**Fig. 17** The spatial distribution of activity centers in the US area of Chicago Twitter users with different detected age groups.

More interestingly, compared to other types of data that often used for human mobility study, the Twitter like location-based social media services provides access to the contents of the social media profiles and messages in addition to the location information. This offers an opportunity to detect the background information (e.g., socioeconomic status, demographic groups and health status) of the social media users, which are critical to understand human mobility and other complex social dynamics. To take advantage of this opportunity, we broke down the Twitter users into different groups at three demographic factors (race/ethnicity, gender and age) by analyzing the names provided in the user profiles. By comparing the human mobility across different demographic groups, we found that the human mobility measures of each group still generally follow the power law distribution, but demonstrate obvious differences across demographic groups. The spatial distribution of home locations and activity centers also demonstrate difference across different groups, particularly between different race/ethnicity groups. Generally, among the three demographic factor of study, the race/ethnicity demonstrates the largest impact on human mobility patterns, then the age, and the gender demonstrates the least.

Findings of this paper lay foundations for future investigations of human mobility and related research. It has been well known that the urban human mobility is refrained by the urban structures and urban morphology. Although this study focuses on the city of Chicago, the analysis, methodology and the work flow can be generalized to other areas with different geographical scales. More importantly, in addition to the three demographic factors mined from the Twitter profile names in this study, other sociodemographic factors, such as occupation and health status, could be mined from the social media contents, and could be included into the analysis of human mobility. Therefore, with an access to the social media contents and user profiles, the location-based social media data provide more potential details that are helpful to further understand the human mobility complex and socioeconomic environment (Liu et al. 2015). As a new modality of data source, the location-based social media data have been used in many different fields applications and sometimes starts to be integrated in the decision making processes. A recognized issue with the location social media data is the bias of social media users in representing the real-world demography distribution (Longley & Adnan 2016) and the low sampling rate (as evidenced in Table 1). Given a geographic area, understanding the demographic and socioeconomic background of the local social media users, as partially addressed in this paper, and comparing it to the census demographic data, becomes critical in evaluating the quality of social media data and the impact on decision making process. In addition to the sampling issue, the location-based social media data, including social media contents and the data mining results of contents, are also subjective to uncertainty. Although this uncertainty issue have been recognized and discussed, the methods and tools that could effectively address the uncertainty issues are still largely lacking. Name analysis and the detection of activity centers (including home locations) play an important role in the analysis of this paper. The definition of an activity center is subjective to the geographic scale at study and specification of threshold count number. In name analysis, to reduce the chance of misclassification, we tend to be conservative by specifying higher threshold values. How to systematically evaluate the performance of this name analysis approach and model the uncertainty warrants further investigation in the future.

## 5 Acknowledgments

We gratefully acknowledge the funding provided by the Office of the Vice President for Research at Texas Tech University and the United States Department of Agriculture (Agricultural Research Services).